\documentclass[a4paper,onecolumn,11pt]{article}
\usepackage[super,numbers, sort&compress ]{natbib}
\usepackage{cancel}
\usepackage[a4paper,
            inner=20mm,
            outer=50mm,
             top=20mm,
            bottom=20mm,
            marginparsep=5mm,
            marginparwidth=65mm,
          ]{geometry}

\usepackage[english]{babel}
\usepackage[utf8]{inputenc}
\usepackage{amsmath}

\usepackage{graphicx}
\usepackage{txfonts}
\usepackage[colorinlistoftodos]{todonotes}
\usepackage{hyperref}
\usepackage{gensymb}
\usepackage{authblk}
\usepackage{pdfpages}

\usepackage{enumitem,amssymb}
\newlist{todolist}{itemize}{2}
\setlist[todolist]{label=$\square$}
\usepackage{pifont}

\usepackage{setspace}

\title{\textbf{Stellar Surface Magnetic Fields Impact Limb Darkening}}

\author[1]{Nadiia~M.~Kostogryz}
\author[1]{Alexander~I.~Shapiro}
\author[1]{Veronika~Witzke}
\author[1]{Robert~H.~Cameron}
\author[1, 2]{Laurent~Gizon} 
\author[1]{Natalie~A.~Krivova}
\author[3]{Hans-G.~Ludwig}
\author[4]{Pierre~F.~L.~Maxted}
\author[5, 6, 7]{Sara~Seager}
\author[1]{Sami~K.~Solanki}
\author[8]{Jeff~Valenti}

\affil[1]{Max-Planck-Institut für Sonnensystemforschung, Justus-von-Liebig-Weg 3, 37077 Göttingen, Germany}
\affil[2]{Institut f\"ur Astrophysik, Georg-August-Universit\"at G\"ottingen, Friedrich-Hund-Platz 1, 37077 G\"ottingen, Germany}
\affil[3]{Zentrum f\"ur Astronomie, Landessternwarte, Königstuhl 12, 69117, Heidelberg, Germany}
\affil[4]{Astrophysics group, Keele University, Keele, Staffordshire ST5 5BG, UK}
\affil[5]{Department of Physics and Kavli Institute for Astrophysics and Space Research, Massachusetts Institute of Technology, Cambridge, MA 02139, USA}
\affil[6]{Department of Earth, Atmospheric and Planetary Sciences, Massachusetts Institute of Technology, Cambridge, MA 02139, USA}
\affil[7]{Department of Aeronautics and Astronautics, Massachusetts Institute of Technology, 77 Massachusetts Avenue, Cambridge, MA 02139, USA}
\affil[8]{Space Telescope Science Institute, 3700 San Martin Drive, Baltimore, MD 21218, USA}

\begin{document}

\maketitle



{\bf 
Stars appear darker at their limbs than at their disk centers because at the limb we are viewing the higher and cooler layers of stellar photospheres. 
Limb darkening derived from state-of-the-art stellar atmosphere models systematically fails to reproduce recent transiting exoplanet light curves from the Kepler, TESS, and JWST telescopes --- stellar brightness obtained from measurements drops less steeply towards the limb than predicted by models. All previous models assumed atmosphere devoid of magnetic fields. Here we use our new stellar atmosphere models computed with the 3D radiative magneto-hydrodynamic code MURaM to show that small-scale concentration of magnetic fields on the stellar surface affect limb darkening at a level that allows us to explain the observations. Our findings provide a way forward to improve the determination of exoplanet radii and especially the transmission spectroscopy analysis for transiting planets, which relies on a very accurate description of stellar limb darkening from the visible through the infrared. Furthermore, our findings imply that limb darkening allows measuring the small-scale magnetic field on stars with transiting planets. }

Efforts to compute stellar limb darkening go back over a century to the classical work of Milne \cite{Milne_1921}. The knowledge of limb darkening is required for numerous astrophysical applications, e.g. measurements of stellar diameters with interferometry\cite{Hestroffer_1997}, the interpretation of light curves of  eclipsing binary stars \cite{Kopal_1950} and spotted stars \cite{Lanza2003}.


The iconic modern-day application of limb darkening is for transit light curve fitting to derive planetary radii with transit photometry and atmospheric composition with transmission spectroscopy \cite{seager_sasselov_2000, brown_2001, charbonneau_2002}. It relies on the description of stellar limb darkening, which alters the transit profile and depth. Yet there is a conundrum: stellar atmosphere models indicate a significantly stronger drop of the brightness towards the stellar limb than multiple sensitive observations show. This point is made clear by analyses of Kepler and TESS transit light curves \cite{maxted_2018, maxted_2023}. Further observations include that of Alpha Centauri \cite{kervella_2017} with the Very Large Telescope Interferometer (VLTI)  and, most recently, that of WASP-39 \cite{rustamkulov_2023_JWST} with the \textit{James Webb} Space Telescope (JWST). In particular, Maxted \cite{maxted_2023} compared observed and modeled limb darkening in 33 Kepler and 10 TESS stars with transiting exoplanets. He found that while all theoretically computed limb darkenings, including those based on MPS-ATLAS\cite{kostogryz2022}, ATLAS\cite{Sing_2010, Claret_Bloemen_2011, Neilson_Lester_2013}, PHOENIX\cite{Claret_2018}, STAGGER\cite{maxted_2018}, and MARCS\cite{Morello_2022} model atmospheres are in relatively good agreement with each other, they all indicate a steeper drop of the stellar brightness from the center of a stellar disk towards its limb than observations. 

One response to this mismatch has been to include empirical limb-darkening coefficients for each wavelength bin as free parameter when fitting transit light curves \cite{espinoza_2015}. However, increasing the number of free parameters introduces biases and additional uncertainties in planetary radius determination \cite{knutson_2007, espinoza_2015, kreidberg_2015, espinoza_2016}. 
The problem is raised to a new level with the advent of the JWST \cite{gardner_2006}. 
The extreme precision transmission spectroscopy data obtained by this telescope require a more accurate transit analysis, which in turn calls for improved  theoretical modeling of limb darkening. Similar precision is also expected from the Extremely Large Telescope (ELT, first light expected in 2027) and ARIEL \cite{tinetti_2018} (launch expected in 2029).

Here we show that stellar surface magnetic fields measurably affect limb darkening and the lack of magnetic field in the theoretical stellar atmosphere models is the main culprit behind the mismatch with observations. All stars on the lower main sequence are intrinsically magnetic \cite{Reiners_2012LRSP}, with the magnetic field being formed within a star and emerging on the surface due to buoyancy \cite{Parker_1955}. Large concentrations of surface magnetic field form active regions (containing, e.g., dark spots). While spots produce an the offset of the transit curve \cite{sag_2022}, they do not change stellar limb darkening unless the planet crosses them during the transit. Smaller concentrations of field lead to the formation of a more homogeneous magnetic network present all over the star so that a transiting planet always crosses it. We simulate the magnetic network on a star with solar fundamental parameters using the 3D radiative magnetohydrodynamics (MHD) code MURaM \cite{Voegler_2005} and radiative transfer code MPS-ATLAS \cite{mps-atlas_2021} and show that the network modifies the limb darkening, making the stellar limb brighter relative to the non-magnetic case. Most importantly, our calculations reconcile models with Kepler \cite{borucki_2010} and TESS \cite{ricker_2015} measurements of limb darkening on stars with near-solar fundamental parameters.

\section*{Results}
Small-scale concentration of magnetic fields change the structure of the stellar photosphere and, consequently, stellar limb darkening. Some of these magnetic fields are generated deep within the stellar interior by the action of a global stellar dynamo \cite{charbonneau_2013}. They affect the structure of the stellar photosphere when they emerge at the stellar surface. Another important component of the stellar surface magnetic field results from the action of a near-surface small-scale turbulent dynamo (SSD) \cite{Rempel2014}. Such an SSD fills nearly the entire stellar surface with considerable magnetic flux, thus producing a minimum level of magnetic activity \cite{Trujillo2004Natur, 2018ApJ...863..164D, bhatia_2022, witzke2022_SSD}. The turbulent magnetic fields produced by SSD are always present at the stellar surface independently of the action of the global stellar dynamo and they also modify the stellar photospheric structure \cite{bhatia_2022, witzke2022_SSD} (relative to the hypothetical non-magnetic case). In this study we simulate the effect of both components of the stellar magnetic field (representing those by global and by small-scale dynamos) on the stellar photosphere with the MURaM code and then utilize the MPS-ATLAS code to calculate the limb darkening. While MURaM is capable of self-consistently simulating the action of an SSD \cite{Rempel2014, Rempel2018, bhatia_2022, witzke2022_SSD}, the effect from magnetic fields brought about by the global dynamo is simulated by adding homogeneous, vertical magnetic fields of 100~G, 200~G, 300~G to the initial SSD setup \cite{witzke_2022} (hereafter, referred to as 100~G, 200~G, 300~G cases, respectively). These initially homogeneous and vertical magnetic fields evolve to a statistically steady state as our simulations relax. This state is highly inhomogeneous: strong, nearly vertical magnetic fields condense in the downflow lanes (integranular lanes) while convective cells (granules) harbor relatively weak, mainly horizontal fields. 

To illustrate our results we first introduce a formalism for deriving the steepness of limb darkening and for quantifying the offset between models and observations. Then we proceed with showing that these offsets can be explained by the magnetic field. We follow the description of limb darkening established in previous studies \cite{maxted_2018, maxted_2023} and define two parameters that characterize its steepness:
\begin{equation*}
h_1'= \frac{I_{\mu=2/3}}{I_{\mu=1}}
\end{equation*}
and
\begin{equation*} 
h_2' = \frac{I_{\mu=2/3}}{I_{\mu=1}} - \frac{I_{\mu=1/3}}{I_{\mu=1}}. 
\end{equation*}
Here $I_{\mu}$ is the intensity in the Kepler or TESS passbands at a specified $\mu$ value, which is the cosine of the heliocentric angle, i.e. the angle between the direction normal to the stellar surface and the direction to the observer. 



Following a previously proposed approach \cite{maxted_2023} we quantify the difference between observations and models by considering the offset between $h_1'$ obtained from observations and from models and the same for $h_2'$ (these offsets are referred to as $\Delta h_1'$ and $\Delta h_2'$). These quantities can obviously also be used to distinguish between the limb darkenings obtained from different sets of models. Here,  $\Delta h_1'$ and $\Delta h_2'$ are first determined for non-magnetic models for which we use set~1 MPS-ATLAS limb darkening taken from Kostogryz el al. \citep{kostogryz2022} (hereafter REFLD), and the observations are the Kepler and TESS samples from Maxted \cite{maxted_2023}. The REFLD accounts for the dependence of the non-magnetic limb darkening on stellar fundamental parameters (effective temperature, metallicity, and surface gravity). Following that we determine  $\Delta h_1'$ and $\Delta h_2'$ for the limb darkening obtained from simulations including magnetic fields. We use solar MURaM simulations to calculate the magnetic effect on limb darkening. 
It is appropriate for the purposes of the present study to restrict these simulations to a solar atmosphere since stars in the considered Kepler and TESS samples have near-solar fundamental parameters. Also, whereas the effect of the magnetic field per se is a first order effect, its dependence on stellar fundamental parameters (for stars with fundamental parameters not too far from the Sun) is a second order effect and represents only a small correction, as a preliminary analysis of recent MURaM simulations of magnetized atmospheres of stars with different fundamental parameters \cite{bhatia_2022, witzke2022_SSD, Yvonne_MNRAS} suggests. In a forthcoming publication we will extend our investigation to a broader sample of stars.

{\bf Limb darkening  without magnetic fields.} The mean offset between the observations and the REFLD model results over all considered targets\cite{maxted_2023} was found to be $\Delta h_1'=0.6\% \pm 0.2\%$ and $\Delta h_2'=-1.2\% \pm 0.4$ for the Kepler sample as well as $\Delta h_1'=0.4\% \pm 0.3\%$ and $\Delta h_2'=-0.9\% \pm 0.4$ for thee TESS sample. Combining the two samples, points at $\mu=2/3$ (corresponding to about 75\% of the projected distance from center to limb) are  observed to be ~0.5\% brighter than predicted by non-magnetic models. Similarly, points at $\mu=1/3$ (corresponding to about 95\% of the projected distance from center to limb) are observed to be ~1.5\% brighter.

\begin{figure}
    \centering
    \includegraphics[width=0.98\linewidth]{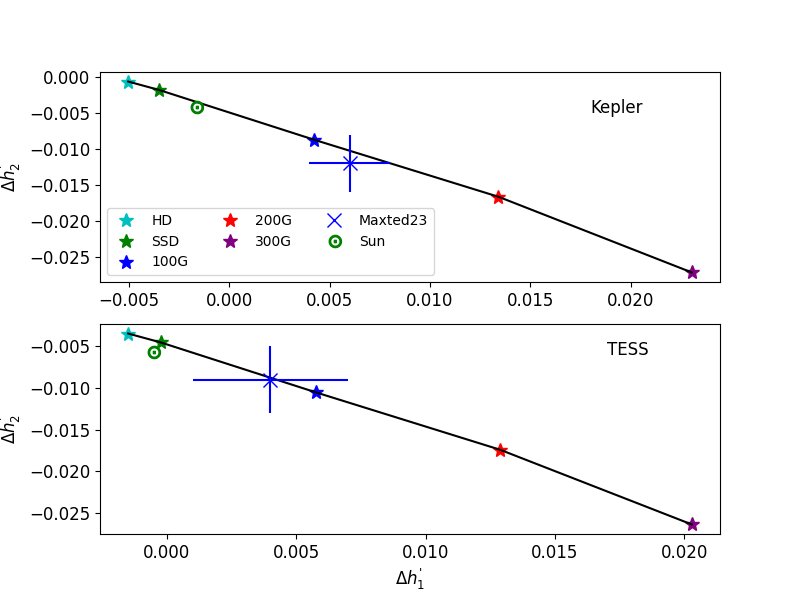}
    \caption{{\bf Limb darkening in Kepler and TESS passbands.} The x- and y-axes express the difference between steepness of limb darkening coefficients ($h_1'$ and $h_2'$) between two sources: one source is always the models without magnetic fields (REFLD), the other can either be models with fields or observations.
    The points marked by an X with error bars are the data (top panel: Kepler, bottom panel: TESS) and show the mean offset between measurements\cite{maxted_2023} and REFLD. The starred points are our calculations based on MURaM simulations (i.e. offsets between our calculations and REFLD) for  different magnetization levels (see legend in upper panel) in Kepler and TESS passbands (upper and lower panel, respectively). The solar sign is the measured solar limb darkening in Kepler and TESS passbands \cite{kostogryz2022}. We note that our zero magnetic field models (HD) are not exactly at the origin because our MURaM calculations without magnetic field (HD) are not identical to REFLD since REFLD uses a simplified treatment of convection. Our stellar atmosphere models, which include magnetic fields, match both, the solar and the stellar observations and provide an explanation for the offsets between TESS and Kepler limb darkening relative to models without magnetic field.}
    \label{fig:delta_h1_h2_to_averaged_obs}
\end{figure}

\begin{figure}
    \centering
    \includegraphics[width=0.98\linewidth]{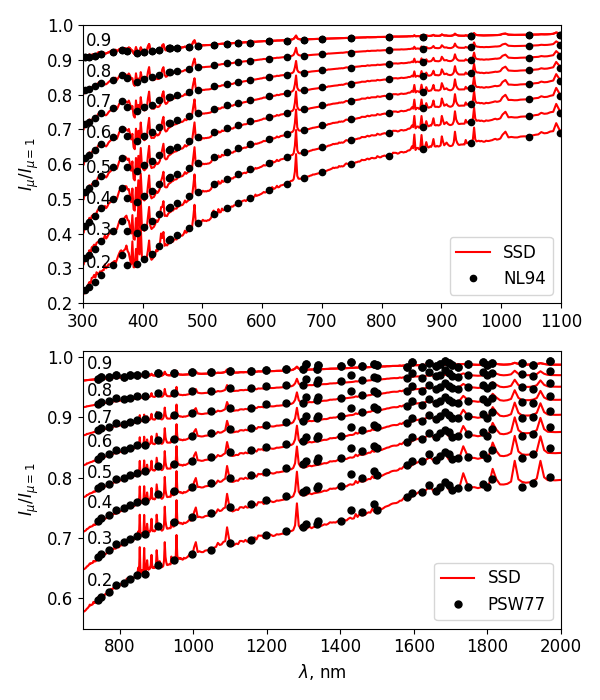}
    \caption{{\bf Solar limb darkening.} Shown are spectral intensities ($I_{\mu}$) at different disk positions given by 
    $\mu$ (labeled on the left side in the figure) normalized to the disk center intensity ($I_{\mu=1}$). Solid lines correspond to our calculations for the SSD-case (i.e. they indicate spectra emerging from MURaM cubes as computed with the MPS-ATLAS code) while  asterisk symbols in the top and bottom panels are solar measurements (by NL94 \cite{NL94} and PSW77 \cite{PSW77}, respectively), where any manifestation of solar magnetic activity are excluded.
    Our model agrees with measurements from UV, through visible to infrared wavelengths.}
    \label{fig:ld_vs_obs}
\end{figure}

{\bf Magnetic limb darkening in Kepler and TESS measurements.} 
The small-scale magnetic fields make the stellar limb darkening 
less steep, leading to an increase of the brightness at $\mu=2/3$  and to an even stronger increase at $\mu=1/3$ (compared to the limb darkening for the quiet star; see Figure~\ref{fig:delta_h1_h2_to_averaged_obs}). The smallest offset relative to the non-magnetic case is caused by the magnetic fields generated by an SSD (green stars in Fig.~\ref{fig:delta_h1_h2_to_averaged_obs}). Despite the fact that an SSD fills the entire solar photosphere with a relatively large magnetic field (e.g., the mean vertical field, $<|B_z|>$, at the visible surface is about 70~G \cite{witzke2022_SSD}), only a small fraction of these fields condense  to local concentrations harboring strong, i.e. kG magnetic fields \cite{Rempel2018, witzke2022_SSD} resulting in only moderate heating of the photosphere. Our results indicate that spatially resolved measurements of the quiet Sun's limb darkening \cite{PSW77, NL94} can be very accurately reproduced by the SSD setup (Figures~\ref{fig:delta_h1_h2_to_averaged_obs}~and~\ref{fig:ld_vs_obs}). This is reassuring since the spatially resolved solar limb darkening observations were performed avoiding visible manifestations of magnetic activity and, thus, should correspond to the minimum possible magnetic activity, which is thought to be produced by the SSD setup (see Methods for more details).

In contrast, to the small-scale magnetic fields generated by SSD, the fields brought about by the global dynamo, e.g. those forming plagues and the magnetic network of the Sun, have a more inhomogeneous spatial distribution and cluster in concentrations of up to several kG \cite{Solanki_et_al_2013, Beeck2015A&Athird}. These concentrations are often described  as magnetic flux tubes \cite{Sami2006} and produce substantial changes in e.g. the thermal structure of the photosphere. Consequently, even a spatially averaged field of 100 G (once it has formed kG concentrations through its interaction with the convection) in addition to the SSD-generated field produces a much stronger change of limb darkening than the pure SSD case. Larger spatially averaged magnetic flux densities, e.g., 200 G and 300 G added to SSD simulations lead to correspondingly larger changes of the limb darkening (Figure~\ref{fig:delta_h1_h2_to_averaged_obs}). 

Our calculations show that Kepler and TESS measurements can be explained by magnetic fields generated by a global dynamo. Indeed, the mean offsets for Kepler and TESS stars are very close to the calculations corresponding to the 100 G case. More importantly, the points representing yhe Kepler and TESS offsets lie very close to the lines connecting points illustrating different magnetic cases (Figure~\ref{fig:delta_h1_h2_to_averaged_obs}).  This implies that our magnetized 3D model atmospheres allows simultaneously explain the offsets in both limb darkening coefficients and, therefore, reproduce the observed limb darkening within the error bars. Clearly, the magnetic field removes the discrepancy in the limb darkenings and its neglect in earlier models is the likely cause of this discrepancy.

We note that our result does not necessarily imply that stars in the Kepler and TESS samples are on average more active than the Sun. While solar limb darkening measurements have been specifically designed to avoid any distortion by magnetic activity, stellar measurements correspond to limb darkening along the transit path without any option to remove contributions from magnetic activity. Consequently, a stronger magnetic field is needed to reconcile stellar models and observations.


\begin{figure*}
    \centering
    \includegraphics[width=\linewidth]{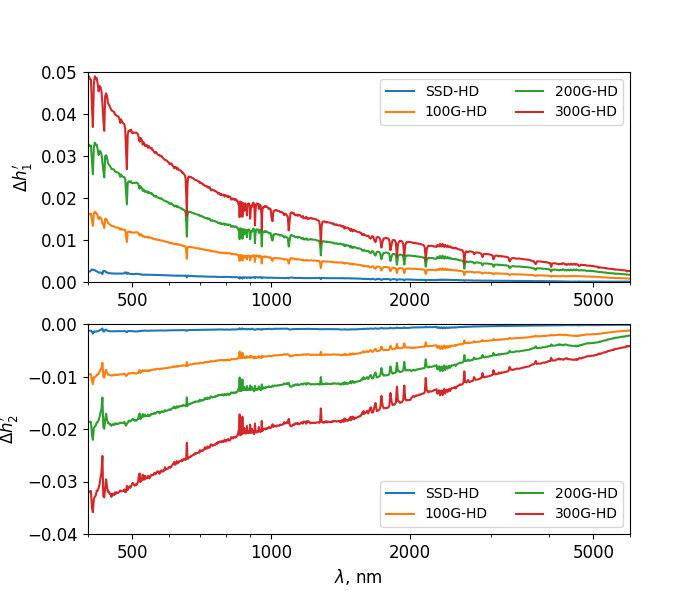}
    \caption{{\bf Wavelength-dependent limb darkening coefficients}. We show the differences between limb darkening coefficients ($\Delta h_1'$ in top panel and $\Delta h_2'$ in bottom panel) resulting from MHD simulations with a magnetic field (SSD, 100~G, 200~G, 300~G) and from non-magnetic (HD simulations). The colors distinguish between the various magnetic (SSD, 100G, 200G, 300G) simulations entering the difference (see legend in the figure). The dotted lines indicate the zero level, e.g. no effect from magnetic fields on stellar limb darkening.  The effect of the surface magnetic field on the limb darkening persists from the UV through the visible and into the infrared spectral domains.}
    \label{fig:ld_coefficients_wavelength}
\end{figure*}

\noindent{\bf Magnetic limb darkening in the JWST era.}
The effect of the magnetic field on limb darkening affects the interpretation of the JWST transmission spectroscopy data. Indeed, the first JWST observations of transits in the WASP-39~b system \cite{rustamkulov_2023_JWST} performed with the NIRSpec PRISM showed that the deviation between observations and currently available models persists over the entire spectral domain of these observations (about 0.5-5.5 $\mu$m). To quantify the effect of magnetic field on limb darkening over the JWST spectral domain we show the dependence of $\Delta h_1'$ and $\Delta h_2'$ on the average magnetic field (with respect to the field-free MURaM HD simulations) as a function of wavelength in Figure~\ref{fig:ld_coefficients_wavelength}.  
The amplitude of the magnetic effect decreases towards longer wavelengths where emergent intensity is less sensitive to temperature changes caused by the magnetic field. At the same time $\Delta h_1'$ and $\Delta h_2'$ have rather complex spectral profiles especially in the visible spectral domain where limb darkening is strongly affected by atomic and molecular lines. 

\begin{figure*}
    \centering
    \includegraphics[width=0.98\linewidth]{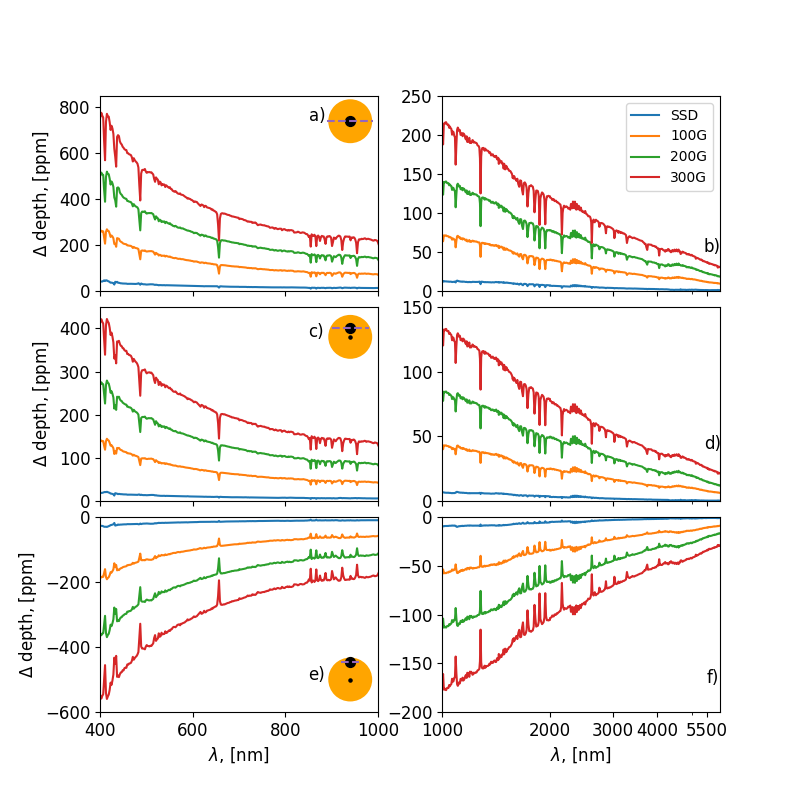}
    \caption{{\bf Effect of surface magnetic field on transit depth for the exemplary case of a Jupiter-size planet transiting a solar twin.} The difference of transit depth ($\Delta$ depth) computed with limb darkening corresponding to different degrees of magnetization (different colors) with respect to non-magnetic limb darkening (HD) in parts per million (ppm). 
    The two columns present the $\Delta$ depth at wavelengths (a, c, e) shorter than and (b, d, f) longer than 1000 nm. The three rows corresponds to different impact parameters ($b=0.0, 0.5, 0.9$, respectively) describing the transit path, which is schematically shown in a, c and e panels. The bigger circle stands for the star and the smaller represents the planet. 
    In general the difference in transit depth is always larger for stronger average stellar fields and is largest in the UV while decreasing towards longer wavelengths.}
    \label{fig:transit_depth}
\end{figure*}

The magnetically-induced change of the limb darkening modifies the entire shape of transit profiles (see Figure~\ref{fig:transitLC} in Methods) and, in particular, the transit depth, which plays a crucial role in determining the radius of transiting planet. The effect is especially strong shortward of 2000 nm, e.g. the magnetic field of 100~G would induce a change of the transit depth larger than about 30--40 ppm for a Jupiter-size planet transiting the Sun (Figure ~\ref{fig:transit_depth}). Such changes can be observed with JWST even for a single transit observation and will also interfere with the interpretation of the JWST transit light curves. Indeed, the first JWST results \cite{rustamkulov_2023_JWST} indicate that the transit curves  are not contaminated by systematic effects in most spectral channels, and the noise can be reduced to just a couple of ppm per spectral bin for most of the exoplanet target stars JWST observes (see Fig.~8 from Rustamkulov et al.\citep{rustamkulov_2023_JWST}).


\section*{Discussion}

Small-scale surface magnetic fields previously ignored in modeling of stellar limb darkening modify the atmospheric structure and therefore stellar limb darkening. We have shown that adding such magnetic fields into 3D radiation MHD simulations of stellar atmospheres solves the limb darkening conundrum, i.e. the inability of 1D and 3D magnetic field-free   models of stellar atmospheres to return limb darkening profiles consistent with observations.

The dependence of the limb darkening on surface magnetic field can be seen as a curse or a blessing depending on one's point of view. On the one hand, it introduces one more free parameter into the light curve fitting. On the other hand, it offers the exciting possibility of measuring the magnetization of stars hosting transiting planets. In particular, the limb darkening method opens the unique opportunity to measure stellar magnetic fields with the upcoming PLATO mission, which 
will observe tens of thousands of bright stars on the lower main sequence \cite{rauer_2014}.

The effect of magnetic field on limb darkening strongly depends on the wavelength (Figs~\ref{fig:ld_coefficients_wavelength}--\ref{fig:transit_depth}) and, thus, ignoring it might introduce spurious features in the transmission spectra obtained with JWST and eventually with ARIEL\cite{tinetti_2018}.  This underscores the importance of accounting for the magnetic effect on limb darkening in the analysis of transmission spectra.  atmospheric retrievals which is offered by our modeling approach.



\section*{Methods}
{\bf Calculations with MURaM and MPS-ATLAS codes.}
We utilize the 3D radiative magnetohydrodynamic (MHD) code MURaM\cite{Voegler_2005, Rempel2014} (which stands for MPS/University
of Chicago Radiative MHD) to simulate the solar atmosphere with different degrees of magnetization using the 'box-in-a-star' approach. Then we use the MPS-ATLAS code \cite{mps-atlas_2021} to synthesize spectra emergent from the MURaM cubes. Both codes have been extensively tested and validated by a number of very sensitive observational tests in numerous publications \cite{Voegler_2005, Shapiro2017, Yeo_et_al_2017, kostogryz2022, witzke_2023}. 

MURaM solves the conservative MHD equations for partially ionized and compressible plasma to model mass, momentum and the energy transport. The transfer of radiative energy is calculated following
a multi-group opacity method \cite{nordlund_1982} using 12 opacity bins \citep{witzke_2023}. The equation of state is determined by utilizing pre-generated look-up tables from the FreeEOS code \cite{Irwin_freeeos_2012} with abundances from Asplund et al\cite{Asplund_2009}. The entropy inflow and pressure at the bottom of the simulated cube are chosen to maintain the effective temperature of 5787 K.

The size of the MURaM box in our simulations is 9~Mm $\times$ 9~Mm (512 $\times$ 512 grid points) in the horizontal direction and 5~Mm (500 grid points) in the vertical direction (4 Mm below the optical surface into the convection zone and 1 Mm above it covering the lower stellar atmosphere and in particular the photosphere). We use the same formulation of the boundary conditions as in Witzke et al.\citep{witzke_2022}. It allows for the deep recirculating of the field through the presence of horizontal field in up-flow regions at the lower boundary of the simulation cube \cite{Rempel2014, Rempel2018}. Such a boundary condition results in the generation of magnetic field at the solar surface whose averaged value does not depend on the depth of the simulation cube \cite{Rempel2014} and allowed explaining ubiquitous small-scale horizontal and mixed polarity magnetic fields that are always present at the solar surface \cite{Danilovic_2010, buehler_etal_2013}. To cover the range of possible facular magnetic flux
densities we also executed simulations with added initially vertical, unipolar and homogeneous magnetic fields of 100 G, 200 G, and 300 G to the setup we used for simulating quiet regions including the SSD. The simulations  are then allowed to relax and the magnetic field to interact with the convection, finally leading to kG magnetic features in the intergranular lanes, separated by regions (granules) with very weak fields. Such an approach allows emulating magnetic fields in facular (plage) and network regions, which are thought to be generated by the action of a global dynamo and, thus, account for their effects on atmospheric structures \cite{Beeck2015A&Athird, Salhab2018, witzke_2022}.

\begin{figure*}
    \centering
    \includegraphics[width=0.98\linewidth]{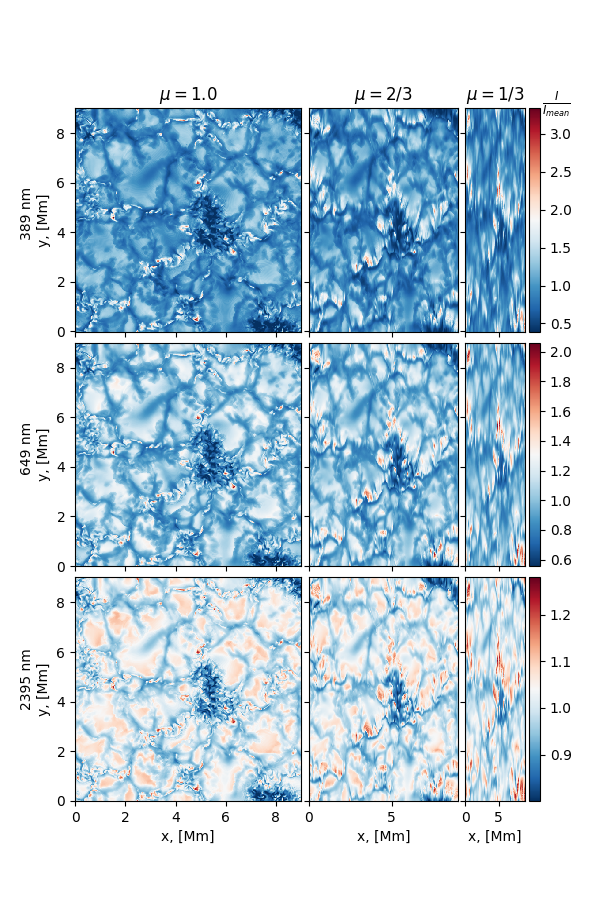}
    \caption{{\bf Still from a movie showing MURaM simulations with an initially imposed vertical field of 300~G.} Shown are values of specific intensity (normalized to the mean intensity in the simulation box)  at three wavelengths (different rows) and at different distances from the limb: at disk center (left column), at $\mu=2/3$ (middle column) and at $\mu=1/3$ (right column). The colorbar on the right of each row describes the normalized intensity at each wavelength. Surface magnetic field condenses in the integranular lanes and leads to the formation of small-scale bright features as well as somewhat larger dark features (small pores; present near the center of the box and at its lower right edge). Both types of features are brighter closer to the limb and, thus, make limb darkening less steep. }
    \label{fig:cubes_wav}
\end{figure*}

We have run our HD and SSD simulations for four solar hours each, while all simulations with added vertical magnetic field have been run for two hours each (after the relaxation). The MPS-ATLAS code \cite{mps-atlas_2021} was then used to calculate spatially-averaged spectra emergent from the cubes with 90-second cadence at 10 disk positions, from disk center ($\mu=1.0$) to the limb ($\mu=0.1$) with a step in $\mu$ of 0.1. The intensity emerging from one snapshot of our 300 G simulations is shown in Fig.~\ref{fig:cubes_wav} at three wavelengths and for three $\mu$ values.

Subsequently, the computed spectra were time-averaged to effectively average out the variability of the spectra caused by granulation and oscillations. Finally, these averaged spectra are used to calculate the limb darkening in the Kepler and TESS passbands.
We have used the time series of our spectra to calculate error of the mean spectrum in the Kepler passband and have checked that it is well below 0.1\% at all disk positions for all considered magnetizations.

{\bf Test against solar measurements.} We compare our calculations of the limb darkening for the quiet solar conditions (SSD case) to spatially-resolved narrow-band solar measurements \cite{PSW77, NL94}  at a number of continuum wavelengths (Figure~\ref{fig:ld_vs_obs}). These measurements carefully selected areas on the solar disk devoid of magnetic activity. Our model demonstrates an excellent agreement with these observations over a wide range of wavelengths. This agreement stems partly from the fact that the 3D MURaM calculations include comprehensive calculations of the convection and overshoot \cite{witzke_2023} which allows a very accurate reproduction of the limb darkening. In particular, our model displays a good performance in the infrared. This is encouraging since most of the transit photometry and spectroscopy measurements are performed in this spectral domain, e.g. mean wavelengths of Kepler and TESS are 630~nm and 800~nm, consequently, while JWST measurements are performed longward 600~nm.

\begin{figure*}
    \centering
    \includegraphics[width=0.98\linewidth]{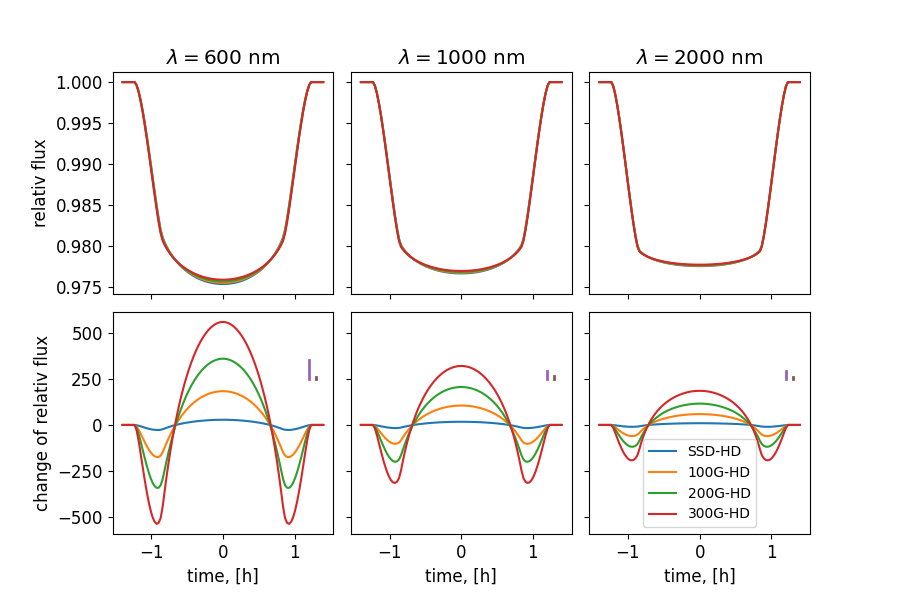}
    \caption{{\bf Simulated WASP-39b transit light curves.} Top panel: the transit light curve at different wavelengths, calculated assuming different levels of WASP-39 atmospheric magnetization. Bottom panels: differences to non-magnetic (HD) calculations. Shown are small-scale dynamo (SSD; blue curve) simulations representing the minimum possible level of magnetic activity as well as fully relaxed simulations with a superposed initial vertical magnetic field of 100 G (orange), 200 G (green), and 300 G (red). The error bars are taken from Rustamkulov et al.\citep{rustamkulov_2023_JWST}.  The larger error bar is the WASP-39b noise from NIRSpec PRISM at the wavelength in our panels (at a bin width of about 0.5\% ). The smaller error bars are for transit curves averaged over 500-nm bins.  The takeaway is that the change in stellar limb darkening due to surface magnetic fields is discernible with the JWST precision. }
    \label{fig:transitLC}
\end{figure*}

{\bf Transit light curves.} The change of the limb darkening induced by surface magnetic field affects the entire transit profile. In Fig.~\ref{fig:transitLC} we illustrate it for the exemplary case of the transit of WASP-39b in front of its host star. The calculations have been performed approximating the limb darkening of WASP-39 by that of the Sun (which is reasonable for illustrative purposes since WASP39 has near-solar fundamental parameters) and using the orbital parameter of WASP-39b from Table 1 of Rustamkulov et al.\citep{rustamkulov_2023_JWST}.

\section*{Data availability}

\section*{Code availability}
The MPS-ATLAS and MURaM codes used in the current study are available on reasonable request.

\bibliographystyle{aa}
\bibliography{refer}

\section*{Acknowledgments}
This work has received funding from the European Research Council (ERC) under the European Union's Horizon 2020 research and innovation program (grant agreement No. 715947). This work has been partially supported from the German Aerospace Center (DLR FKZ~50OP1902). This work was supported in part by ERC Synergy Grant WHOLESUN~810218 and DLR grant PLATO Data Center (DLR FKZ~50OP1902).

\section*{Author information}

Contributions: N.M.K., A.I.S., and S.K.S. conceived the study. N.M.K., V.W., and A.I.S. conducted the simulations and analyzed the results. N.M.K., A.I.S., S.S, S.K.S, and J.V. wrote the manuscript. R.H.C., L.G., N.A.K., H.-G.L., P.F.L.M., S.S., S.K.S., and J.V. contributed to the analysis of the data. All authors discussed the results and reviewed the manuscript.

\noindent{Correspondence to Nadiia Kostogryz}

\section*{Ethic declaration}

The authors declare no competing interests.




\end{document}